# QUANTITATIVE DESCRIPTION AND CORRECTION OF LONGITUDINAL DRIFTS IN THE FERMILAB LINAC*

R. Sharankova[†], A. Shemyakin, S. Rego[1], Fermi National Accelerator Laboratory, Batavia, USA
[1]also at École Polytechnique Palaiseau, France

*Abstract*

The Fermi National Accelerator Laboratory (Fermilab) Linac [1] accepts 750 keV H- ions from the front end and accelerates them to 400 MeV for injection into the Booster rapid cycling synchrotron. Day-to-day drifts in the beam longitudinal trajectory during regular operation are of the order of several degrees. They are believed to cause additional losses in both the Linac and the Booster and are addressed by empirically adjusting cavity phases of front end and Linac RF cavities. This work explores a scheme for expressing these drifts in terms of phase shifts in the low-energy part of the Linac. Such a description allows for a simplified visual representation of the drifts, suggests a clear algorithm for their compensation, and provides a tool for estimating efficiency of such compensation.

## THE FERMILAB LINAC

The front end comprises an Ion Source, a Low Energy Beam Transport (LEBT) line, a radio-frequency quadrupole (RFQ), and a Medium Energy Transport (MEBT) line. The magnetron ion source produces 35 keV H- ions which are chopped, bunched and accelerated to 750 keV of kinetic energy before entering the Linac. The Linac comprises three parts: a Drift Tube Linac (DTL), a transition section and a Side Coupled Linac (SCL). The DTL is composed of 207 drift tubes spread across 5 tanks. It operates at RF frequency of 201.25 MHz and accelerates beam to 116.5 MeV. The SCL has 448 cells across 7 modules and operates at a resonant frequency of 805 MHz, accelerating beam to 401.5 MeV. Longitudinal matching between the DTL and SCL is done by a "buncher" and a "vernier" side-coupled cavities. During regular operations, the Fermilab Linac has an output of roughly 25 mA and pulse length of 35 µs, with transition efficiency of the order of 90%.

## LONGITUDINAL DRIFTS

The beam longitudinal trajectory, reconstructed from phase measurements of Beam Position Monitors (BPMs) [2] along the Linac, drifts day-to-day by order of several degrees. This can be largely attributed to cavity resonance and RF component phase fluctuations due to temperature and humidity changes. Another contributing factor are variations in the phase-space of the beam emerging from the front end. Figure 1 shows the evolution of the beam longitudinal trajectory over a period of 10 hours during regular beam operation, without intentional changes to RF parameters.

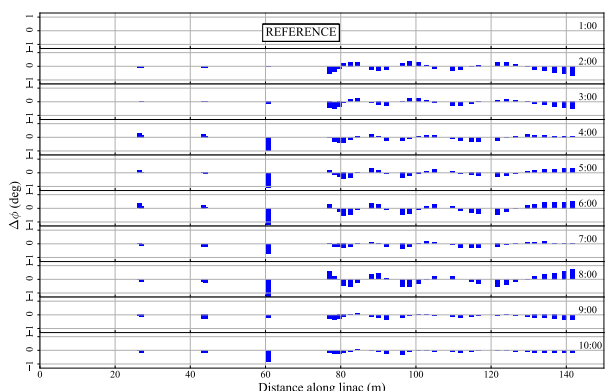

Figure 1: Longitudinal trajectory drift over 10 hours of regular operation.

The horizontal slices in Figure 1 each show data taken at the beginning of a one-hour period, and the first one (corresponding to 1:00 AM) is taken as the reference. Each vertical bar represents the beam phase difference w.r.t. the reference measured by one BPM. The horizontal axis shows the longitudinal coordinate of BPMs along the Linac. The maximum difference of measured phase w.r.t. the reference varies by more than 2 degrees at 402.5 MHz over the 10-hour period. Such drifts are known to affect beam losses in the Linac as well as the Booster at injection. Historically, the Linac has been tuned empirically by changing the RF cavity phases of a handful of cavities in the front end and the DTL. Typically, such tuning aims to minimize overall loss and maximize Linac throughput, and not necessarily to compensate for longitudinal trajectory drifts. Below we describe a method for estimating and compensating longitudinal drifts with the aim of stabilizing the trajectory.

## DESCRIPTION OF LONGITUDINAL DRIFTS

It can be seen from Figure 1 that the density of BPMs is very uneven along the length of the Linac. The DTL is instrumented with only 5 BPMs while the SCL has 28. Simulation shows ~6 synchrotron oscillations in the DTL and ~3 in the SCL. This means that the available instrumentation is only sufficient to describe the beam longitudinal trajectory in the SCL and not the DTL. Having this in mind, the authors developed a scheme for characterizing the longitudinal drifts in the SCL alone. The method describes these drifts as a linear combination of perturbations resulting from shifts in the phase of two cavities upstream of the SCL. In terms of beam phase-space (E,φ), one can think that the sum of two orthogonal vectors with variable amplitudes can represent any arbitrary vector.

*Longitudinal Response Matrix*

To quantify the effect of cavity phase shifts on the beam phase along the SCL, we measure the BPM differential trajectory as a function of cavity phase setting for RFQ, Buncher and DTL Tanks 1-5. Figure 2 shows the result. The horizontal axis corresponds to BPM longitudinal coordinates. The vertical axis is the phase difference measured by each BPM for 1º change in the corresponding cavity phase. In other words, each curve represents a normalized response trajectory for a particular cavity. The combination of all response trajectories is referred to as the longitudinal response matrix.

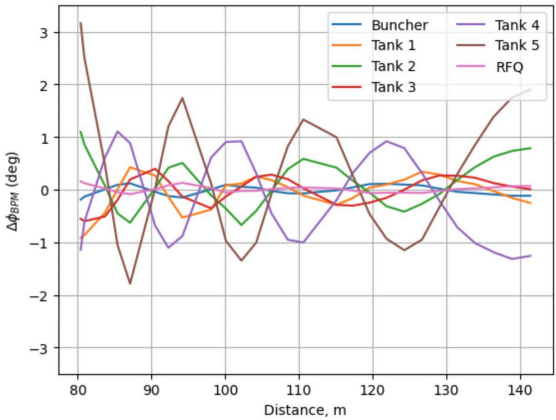

Figure 2: BPM response to 1 deg change in cavity phase for front end and DTL cavities.

*Data Collection with Parameter Oscillation*

The data for BPM response matrix generation was collected utilizing a technique that allows for improved signal-to-noise ratio [3]. Instead of doing a flat scan of cavity phase, we oscillate the cavity phase setting in a sinusoidal pattern at a fixed frequency. Then the BPM data is analysed with Fourier transform, and only the data corresponding to the cavity phase driving frequency is selected. The flat off-frequency component corresponds to the random noise.

This technique also allows us to scan multiple cavity phases at once by driving them at different frequencies that separate the response in the frequency domain. Finally, since the magnitude of the signal peak in the frequency domain can be enhanced by increased the number of measurements instead of the larger amplitude of oscillation, this method is well-suited for collecting data in a parasitic way during regular beam operation.

Figure 3 shows an example of how cavity phases are oscillated to collect response matrix data. The horizontal axis represents number of beam pulses (measurements) while the vertical axes show phase oscillation magnitude relative to the nominal cavity phase in degrees.

Figure 4 shows the Fourier transform of the response of one SCL BPM, located just upstream of Module 1, to the cavity phase oscillation shown in Figure 3.

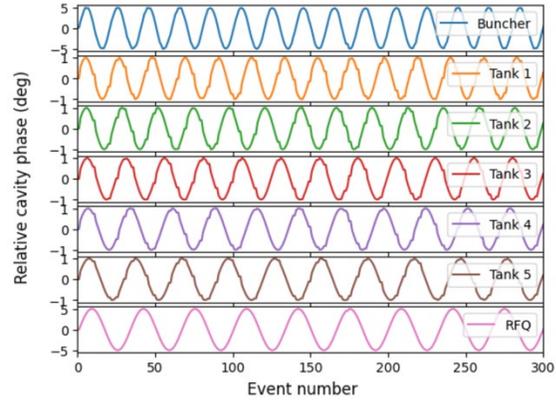

Figure 3: Cavity phase oscillation at different frequencies.

The horizontal axis shows frequency in number of Fourier harmonics while the vertical axis shown BPM phase change in degrees. The peaks are aligned with the driving frequencies of each cavity. The figure inset shows the BPM response in the time domain.

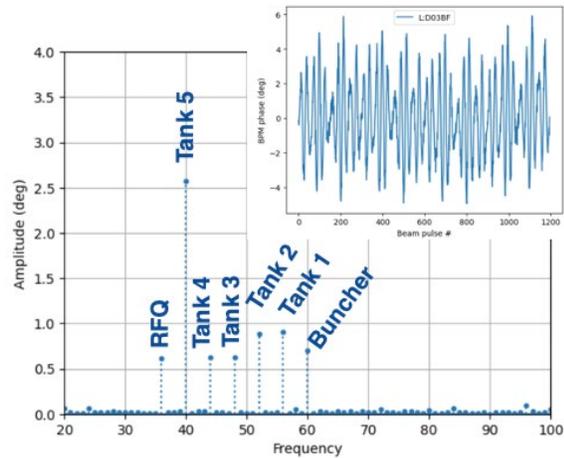

Figure 4: BPM D03 reponse to cavity oscillation in Fourier (main) and time (inset) domain.

## LONGITUDINAL COMPENSATION

*Linear Fitting of Trajectories*

To describe an arbitrary longitudinal trajectory (such as a longitudinal drift caused by a perturbation upstream of the SCL), it is fitted to a linear combination of two of the vectors of the response matrix. Specifically, the two that are closest to orthogonality are selected by calculating the inner product of all response vectors:

$$\cos \alpha_{ij} = \vec{T_i} \cdot \vec{T_j} / (|\vec{T_i}||\vec{T_j}|),$$

where $i,j$ are indices that run over RFQ, Buncher, DTL Tanks 1-5, and T is the response vector (trajectory) for that cavity as shown in Figure 2. The vectors closest to $\alpha_{ij} = \pi/2$ are selected. For the response vectors shown in

Figure 2, the vectors for Tank 3 and Tank 4 satisfy this condition. Next, the target trajectory is fitted using least mean squares method.

The method was tested on data by deliberately shifting one of the upstream cavities and fitting the resulting longitudinal trajectory. Figure 5 shows the result of fitting to data with Buncher cavity phase shifted by +20º. The fit residuals are of sub-percent levels.

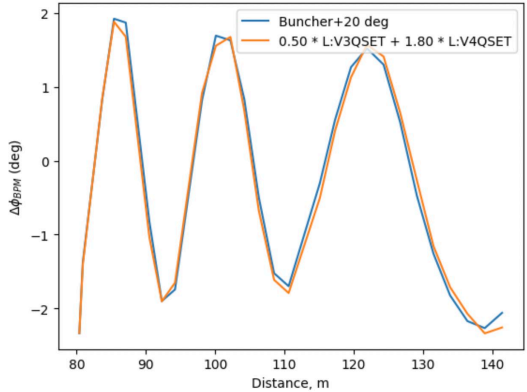

Figure 5: Test of linear fitting of a longitudinal trajectory.

*One-Step Longitudinal Compensation*

The simplest procedure to compensate a longitudinal drift originating upstream of the SCL includes the following steps:

1. Record cavity phase oscillation data and generate response matrix (can be omitted if previous measurement is close in time).
2. Select response vectors closest to orthogonal – linear fit basis vectors.
3. Determine linear fit coefficients for the basis vectors.
4. Apply changes to basis cavity phases by inverting fit coefficient sign.

The procedure was used to correct the daily drift in beam phase on two days in February (15$^{th}$ and 21$^{st}$), with February 8$^{th}$ used as reference. The original longitudinal trajectories and the compensated residuals are shown in Figure 6. Two metrics are calculated to evaluate the effectiveness of the compensation: the mean and the standard deviation (STD) of each trajectory. If the compensation is 100% effective the mean and STD will be approaching 0. The results are summarized in Table 1.

Table 1: One-Step Compensation Results

|  | Mean (deg) | STD (deg) |
|---|---|---|
| Feb 15$^{th}$ | 0.350 | 0.576 |
| Feb 15$^{th}$ comp. | 0.292 | 0.311 |
| Feb 21$^{st}$ | -0.007 | 1.366 |
| Feb 21$^{st}$ comp. | 0.212 | 0.508 |

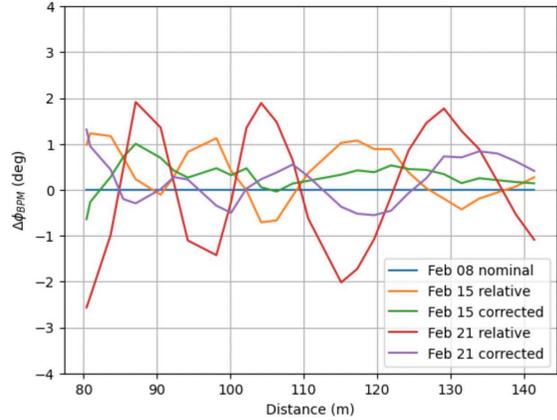

Figure 6: Test of one-step compensation of longitudinal drift on two days in February 2024.

The compensation is O(50%) effective. This can be attributed to drifts of BPM signals and cavity phase set point and stability within the SCL as well as the DTL. The latter can change the reference trajectories themselves.

*Multi-Step Compensation*

To account for uncertainties in the corrections, another procedure was devised which repeats steps (3) and a modified (4) where the compensation coefficients are applied with a factor of 0.5. The steps are repeated until the residual trajectory converges. An example is shown in Figure 7 where a shift in Buncher phase by 20º is corrected in four steps. The method appears to work well and will be tested on daily drift data.

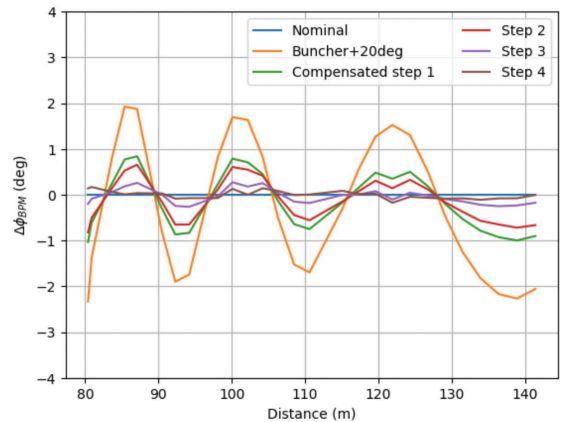

Figure 7: Example of multi-step compensation.

## CONCLUSION

A method was developed to describe longitudinal trajectory drifts in the Fermilab SCL as a linear combination of response vectors to shifts in front end and DTL cavity phases. A compensation strategy was developed based on inverting fit coefficients to these vectors, both in a single and in multiple steps to improve convergence.